\documentclass[12pt,a4]{article}
\pdfoutput=1
\usepackage{latexsym}
\usepackage{mathrsfs}
\usepackage{amssymb}
\usepackage{amsmath}
\usepackage{graphicx}
\usepackage{braket}
\usepackage{enumerate}
\usepackage[usenames,dvipsnames]{xcolor}
\usepackage{todonotes}

\newcommand{\C}{\mathbb{C}}

\newcommand{\SL}{\mathrm{SL}}

\newcommand{\vol}{\mathrm{Vol}}

\newcommand{\rd}{\, \mathrm{d}}

\newcommand{\be}{\begin{equation}\label}
\newcommand{\ee}{\end{equation}}
\newcommand{\bea}{\begin{eqnarray}\label}
\newcommand{\eea}{\end{eqnarray}}

\begin{document}

\title{Soft Theorems from Conformal Field Theory}
\author{Arthur E. Lipstein \vspace{7pt}\\ \normalsize \textit{
II. Institute for Theoretical Physics, University of Hamburg,}\\\normalsize\textit{
Luruper Chaussee 149
22761 Hamburg, Germany}}
\maketitle
\begin{abstract}
Strominger and collaborators recently proposed that soft theorems for gauge and gravity amplitudes can be interpreted as Ward identities of a 2d CFT at null infinity. In this paper, we will consider a specific realization of this CFT known as ambitwistor string theory, which describes 4d Yang-Mills and gravity with any amount of supersymmetry. Using 4d ambtwistor string theory, we derive soft theorems in the form of an infinite series in the soft momentum which are valid to subleading order in gauge theory and sub-subleading order in gravity.  Furthermore, we describe how the algebra of soft limits can be encoded in the braiding of soft vertex operators on the worldsheet and point out a simple relation between soft gluon and soft graviton vertex operators which suggests an interesting connection to color-kinematics duality. Finally, by considering ambitwistor string theory on a genus one worldsheet, we compute the 1-loop correction to the subleading soft graviton theorem due to infrared divergences.
\end{abstract}

\pagebreak
\tableofcontents

\section{Introduction}

In 1965, Weinberg showed that tree-level scattering amplitudes behave in a universal way when the energy of an external photon or graviton goes to zero and the amplitudes are expanded to leading order in the soft momentum \cite{Weinberg:1965nx}. This is known as Weinberg's soft theorem for photons and gravitons. Since then, the soft theorems have been generalized to subleading order for gluons \cite{Low:1958sn,Burnett:1967km,Casali:2014xpa} and sub-subleading order for gravitons \cite{White:2011yy,Cachazo:2014fwa}. Moreover, loop corrections to the soft theorems were studied in \cite{Bern:2014oka,Cachazo:2014dia,He:2014bga,Larkoski:2014bxa}, and double soft limits have been studied in \cite{ArkaniHamed:2008gz,Chen:2014xoa,Cachazo:2015ksa,Klose:2015xoa,Volovich:2015yoa}.

In a recent series of papers, Strominger and collaborators argued that the soft theorems can be interpreted as Ward identites for certain asymptotic symmetries. In particular, they argued that the leading and subleading soft graviton theorems are associated with spontaneously broken BMS symmetries \cite{Strominger:2013jfa,He:2014laa,Kapec:2014opa}, which preserve the conformal structure of null infinity in four dimensional asymptotically flat spacetimes \cite{Sachs:1962wk, Bondi:1962px,Barnich:2009se}, and the soft photon theorems are associated with spontaneouly gauge symmetry at null infinity \cite{He:2014cra,Lysov:2014csa,Kapec:2014zla}. More recently, it has been suggested that the soft gluon theorems can be interpreted as Ward identities associated with a Kac-Moody algebra of a CFT at null infinity \cite{Strominger:2013lka,He:2015zea}. 

In this paper, we will study the soft theorems from the perspective of a 2d CFT which describes 4d Yang-Mills and gravity with any amount of supersymmetry, known as 4d ambitwistor string theory \cite{Geyer:2014fka}. Ambitwistor string theories in general dimensions were developed by Mason and Skinner in \cite{Mason:2013sva}. The key features of these models are that their spectra only contain field theory degrees of freedom and their correlation functions produce scattering amplitudes in the form discovered by Cachazo, He, and Yuan (CHY) \cite{Cachazo:2013gna,Cachazo:2013hca,Cachazo:2013iea,Cachazo:2014xea}, notably they are expressed as worldsheet integrals which localize onto solutions of the scattering equations. The soft theorems were proven in any dimension using the CHY formulae in \cite{Cachazo:2013hca,Schwab:2014xua,Zlotnikov:2014sva,Kalousios:2014uva}, as well as using ambitwistor string theory in \cite{Geyer:2014lca}. They were also proven for $\mathcal{N}=8$ supergravity \cite{Adamo:2014yya} and more recently $\mathcal{N}=4$ super-Yang-Mills \cite{Adamo:2015fwa} using models which are closely related to the 4d ambitwistor string.  

As demonstrated in \cite{Geyer:2014lca}, the soft theorems for gauge theory and gravity can be derived using ambitwistor string theory by expanding a soft vertex operator in powers of the soft momentum. In particular, each term in the expansion corresponds to a charge on the worldsheet which gives rise to soft theorems when inserted into correlation functions. Hence, the soft theorems can be interpreted as Ward identities of ambitwistor string theory. In this paper, we show that if one evaluates the Ward identities in the approximation that the  scattering equations for the hard vertex operators decouple from those of the soft vertex operator, one obtains soft theorems in the form of an infinite series in the soft momentum which are valid to subleading order in Yang-Mills theory and sub-subleading order in gravity. This derivation demonstrates the universality of the soft theorems and clarifies the approximations used in \cite{Geyer:2014lca}. In principle, it should be possible to compute higher order terms in the soft limit by taking into account the backreaction of the soft vertex operator on the hard ones via the scattering equations. Note however, that higher order terms vanish in the holomorphic soft limit, which corresponds to expressing the soft momentum in bispinor form and taking one spinor to zero while holding the other one fixed \cite{Cachazo:2014fwa}. Furthermore, we find an interesting relation between the charges which generate soft gluon and soft graviton theorems. In particular, the latter can be obtained from the former by replacing a Kac-Moody current with a Lorentz generator, which is reminiscent of color-kinematics duality \cite{Bern:2008qj,Monteiro:2011pc}. 

We also show that algebra of soft limits of scattering amplitudes can be elegantly encoded in the braiding of soft vertex operators on the worldsheet of the ambitwistor string. From this point of view, an ambiguity which arises in the definition of double soft limits recently discussed in \cite{He:2015zea} is related to terms which arise from braiding one soft vertex operator around another one before it becomes soft, and our prescription will be to discard such terms. Using this prescription, we find that the commutator of leading order soft graviton limits vanishes, which is what one expects since the underlying symmetry corresponds to supertranslations, which are abelian. Including higher order terms in the soft limit leads to nonzero commutators, which indicates that the symmetry algebra underlying higher order soft theorems is nonabelian. 

Finally, we compute the 1-loop correction to the subleading soft graviton theorem by considering the ambitwistor string on a genus one worldsheet. At leading order, the soft graviton theorem is not renormalized, but at subleading order it receives corrections from 1-loop IR divergences \cite{Weinberg:1965nx,Bern:2014oka}. One-loop amplitudes in ambitwistor string theory were first studied in \cite{Adamo:2013tsa}, where they were shown to have support on the genus one scattering equations. Demonstrating that ambitwistor string theory computes field theoretic loop amplitudes is a difficult task, although this was verified in the IR limit of the 4-point 1-loop integrand in \cite{Casali:2014hfa}. In this paper, we show that in the IR limit, the ambitwistor string loop integrand is a rational function for any number of external legs and can be integrated using dimensional regularization to obtain the 1-loop correction to the subleading soft graviton theorem coming from IR divergences. This provides further evidence that the loop amplitudes of ambitwistor string theory correspond to field theory amplitudes.

This paper is organized as follows. In Section \ref{review}, we review 4d ambitwistor string theory. In Section \ref{tree}, we derive the tree-level soft theorems for Yang-Mills and gravity and compare our expressions to the results of BCFW recursion, from which we deduce that our formulae are valid to subleading order for Yang-Mills and sub-subleading order for gravity. In Section \ref{alsy}, we describe a simple relation between soft gluon and soft graviton vertex operators which is reminscent of color-kinematics duality and explain how the algebra of soft limits can be encoded in the braiding of soft vertex operators on the worldsheet. In Section \ref{loops}, we review ambitwistor string theory in general dimensions both at genus zero and genus one, and compute the 1-loop IR divergent correction to the subleading soft graviton theorem. In Appendix \ref{4dscatt}, we describe some basic properties of the 4d scattering equations. In Appendix \ref{details}, we demonstrate how the soft limit algebra can be encoded in the braiding of soft vertex operators using explicit examples and compute the subleading contribution to the commutator of two soft graviton limits and two soft photon limits (note that the leading contribution vanishes in both cases).               

\section{Review of 4d Ambitwistor Strings} \label{review}

4d ambitwistor string theories were studied in \cite{Bandos:2014lja,Geyer:2014fka}. They are closely related to the twistor string theories of Witten \cite{Witten:2003nn}, Berkovits \cite{Berkovits:2004hg}, and Skinner \cite{Skinner:2013xp}. The crucial  new features of 4d ambitwistor strings are that the worldsheet fields have conformal weight $\left(\frac{1}{2},0\right)$, and vertex operators are defined for both positive and negative helicity particles. Ultimately, this makes 4d ambitwistor strings a lot more flexible than ordinary twistor strings, in that they can describe 4d Yang-Mills theory and gravity with any amount of supersymmetry, and the resulting amplitudes are much simpler, depending on very few moduli.  Whereas super-Yang-Mills theory is described by a non-supersymmetric worldsheet theory, supergravity is described by a worldsheet theory with $\mathcal{N}=2$ supersymmetry. Let us briefly
review this construction. 

The non-supersymmetric worldsheet theory has fields
\[
Z^A=\left(\begin{array}{c}
\lambda_{\alpha}\\
\mu^{\dot{\alpha}}\\
\chi^{a}
\end{array}\right),\,\,\, W_A=\left(\begin{array}{c}
\tilde{\mu}^{\alpha}\\
\tilde{\lambda}_{\dot{\alpha}}\\
\tilde{\chi}_{a}
\end{array}\right)
\]
where $\alpha,\dot{\alpha}$ are spinor indices which are raised and lowered using the two-index Levi-Civita symbol, and the number of fermions $\chi,\tilde{\chi}$ depends on the amount
of target space supersymmetry. We use the following notation to denote spinor inner products: $\left\langle rs\right\rangle =r_{\alpha}s_{\beta}\epsilon^{\alpha\beta}$ and $\left[rs\right]=r^{\dot{\alpha}}s^{\dot{\beta}}\epsilon_{\dot{\alpha}\dot{\beta}}$. The Lagrangian is 

\begin{equation}
\mathcal{L}=W_{A}\bar{\partial}Z^{A}+uW_{A}Z^{A}\label{eq:ly}
\end{equation}
where $u$ is a $GL(1)$ gauge field. The fields have the following
OPE's: 

\begin{equation}
\lambda_{\alpha}(\sigma)\tilde{\mu}^{\beta}(\sigma')\sim\frac{\delta_{\alpha}^{\beta}}{\sigma-\sigma'},\,\,\,\mu^{\dot{\alpha}}(\sigma)\tilde{\lambda}_{\dot{\beta}}(\sigma')\sim\frac{\delta_{\dot{\beta}}^{\dot{\alpha}}}{\sigma-\sigma'}.\label{eq:ope}
\end{equation}
The spectrum of this model contains both (super)Yang-Mills theory and conformal (super)gravity. The integrated
vertex operator for a positive helicity gluon with supermomentum $\left(\lambda_{i}\tilde{\lambda}_{i},\lambda_{i}\tilde{\eta}_{i}\right)$
is 

\begin{equation}
\mathcal{V}_{YM}=\int\frac{\rd t}{t}\delta^{2}\left(t\lambda-\lambda_{i}\right)e^{it\left(\left[\mu\tilde{\lambda}_{i}\right]+\left[\chi\tilde{\eta_{i}}\right]\right)}j\label{eq:ymvo},
\end{equation}
where $j$ obeys a $U(N)$ Kac-Moody algebra

\begin{equation}
j^{A}(\sigma)j^{B}(\sigma')=\frac{k\delta^{AB}}{2\left(\sigma-\sigma'\right)^{2}}+\frac{i f_{\,\,\,\,\,\,\,\,\, C}^{AB} j^{C}}{\sigma-\sigma'}.\label{eq:ca}
\end{equation}
The integrated vertex operator for a negative helicity gluon is
the complex conjugate of \eqref{eq:ymvo}.

The $\mathcal{N}=2$ worldsheet theory has the following additional
world-sheet fields:

\[
\rho^A=\left(\begin{array}{c}
\rho_{\alpha}\\
\rho^{\dot{\alpha}}\\
\omega^{a}
\end{array}\right),\,\,\,\tilde{\rho}_A=\left(\begin{array}{c}
\tilde{\rho}^{\alpha}\\
\tilde{\rho}_{\dot{\alpha}}\\
\tilde{\omega}_{a}
\end{array}\right)
\]
which are the superpartners of $(Z,W)$. The Lagrangian is 

\begin{equation}
\mathcal{L}=W_{A}\bar{\partial}Z^{A}+\tilde{\rho}_{A}\bar{\partial}\rho^{A}+u^{B}K_{B}\label{eq:lg}
\end{equation}
where 

\[
K_{B}=\left\{ W_{A}Z^{A},\tilde{\rho}_{A}\rho^{A},\rho^{\alpha}\rho_{\alpha},\tilde{\rho}^{\dot{\alpha}}\tilde{\rho}_{\dot{\alpha}},\rho^{A}W_{A},Z^{A}\tilde{\rho}_{A},\lambda^{\alpha}\rho_{\alpha},\tilde{\lambda}^{\dot{\alpha}}\tilde{\rho}_{\dot{\alpha}}\right\} .
\]
The superpartner fields have the following OPE's:

\begin{equation}
\rho_{\alpha}(\sigma')\tilde{\rho}^{\beta}(\sigma)\sim\frac{\delta_{\alpha}^{\beta}}{\sigma-\sigma'},\,\,\,\rho^{\dot{\alpha}}(\sigma')\tilde{\rho}_{\dot{\beta}}(\sigma)\sim\frac{\delta_{\dot{\beta}}^{\dot{\alpha}}}{\sigma-\sigma'}.\label{eq:ope2}
\end{equation}
The integrated vertex operator for a positive helicity graviton with supermomentum
$\left(\lambda_{i}\tilde{\lambda}_{i},\lambda_{i}\tilde{\eta}_{i}\right)$
is 
\begin{equation}
\mathcal{V}_{GR}=\int\frac{\rd t}{t^{2}}\delta^{2}\left(t\lambda-\lambda_{i}\right)\left(\left[\tilde{\lambda}\tilde{\lambda}_{i}\right]+it\left[\tilde{\rho}\tilde{\lambda}_{i}\right]\left[\rho\tilde{\lambda}_{i}\right]\right)e^{it\left(\left[\mu\tilde{\lambda}_{i}\right]+\left[\chi\tilde{\eta_{i}}\right]\right)}.
\label{gv}
\end{equation}
The integrated vertex operator for a negative helicity graviton is the complex conjugate of \eqref{gv}.

An N$^{k-2}$MHV amplitude is computed from a correlator with $k$
negative helicity vertex operators and $n-k$ positive helicity vertex
operators: 

\[
\left\langle \tilde{\mathcal{V}}_{1}...\mathcal{\tilde{V}}_{k}\mathcal{V}_{k+1}...\mathcal{V}_{n}\right\rangle. 
\]
For more details about gauge-fixing and BRST invariance, see \cite{Geyer:2014fka}. To simplify the discussion, we will focus on pure Yang-Mills and Einstein gravity for the remainder
of the paper. It is straightforward to generalize our results to the
supersymmetric case. 

Combining the arguments of the exponentials in the vertex operators with the action
and integrating out the worldsheet fields $\mu,\tilde{\mu}$ implies that 

\begin{equation}
\lambda(\sigma)=\sum_{i=1}^{k}\frac{t_{i}\lambda_{i}}{\sigma-\sigma_{i}},\,\,\,\tilde{\lambda}(\sigma)=\sum_{i=k+1}^{n}\frac{t_{i}\tilde{\lambda}_{i}}{\sigma-\sigma_{i}}.
\label{scatts1}
\end{equation}
Plugging these solutions back into the delta functions of the vertex
operators then gives

\begin{equation}
\Pi_{i=1}^{k}\delta^{2}\left(t_{i}\tilde{\lambda}\left(\sigma_{i}\right)-\tilde{\lambda}_{i}\right)\Pi_{j=k+1}^{n}\delta^{2}\left(t_{j}\lambda\left(\sigma_{j}\right)-\lambda_{j}\right).
\label{scatts2}
\end{equation}
These delta functions localize the worldsheet integrals onto solutions
of the 4d scattering equations refined by helicity: 

\[
\left[\tilde{\lambda}_{i}\tilde{\lambda}\left(\sigma_{i}\right)\right]=0,\,\,\, i=1,...,k,\,\,\,\left\langle \lambda_{j}\lambda\left(\sigma_{j}\right)\right\rangle =0,\,\,\, j=k+1,...,n.
\]
Note that these equations describe the scattering amplitudes of both
Yang-Mills theory and gravity. We describe  various properties of these equations in Appendix \ref{4dscatt}.

\section{Tree-Level Soft Theorems} \label{tree}

In this section, we will use ambitwistor string theory to derive soft gluon and graviton theorems in the form of an infinite series in the soft momentum. We then compare our formulae to the results of BCFW recursion and deduce that they are valid up to subleading order in Yang-Mills theory and sub-subleading order in gravity (note that higher order terms vanish in the holomorphic soft limit). This approach is closely related to the one developed in \cite{Geyer:2014lca}, and demonstrates the universality of the soft theorems while clarifying the approximations used in deriving them using ambitwistor string theory.    

\subsection{Yang-Mills}

As explained in \cite{Geyer:2014lca}, from the point of view of ambitwistor string theory, a soft limit corresponds to taking a vertex operator soft in a correlation function. First note that the delta function which appears in a positive helicity vertex operator can be written as

\[
\delta^{2}\left(t\lambda(\sigma)-\lambda_{s}\right)=\delta\left(t-\frac{\left\langle s\xi\right\rangle }{\left\langle \lambda(\sigma)\xi\right\rangle }\right)\delta\left(\left\langle s\lambda(\sigma)\right\rangle \right)
\]
where $\lambda_{s}\tilde{\lambda}_{s}$ is the soft momentum, $\xi$
is a reference spinor, and $s$ is short for $\lambda_s$. Noting that $\bar{\partial}j=0$ and

\[
\delta\left(\left\langle s \lambda(\sigma)\right\rangle \right)=\frac{1}{2\pi i}\bar{\partial}\frac{1}{\left\langle s \lambda(\sigma)\right\rangle },
\]
Stokes theorem implies that

\[
\int \rd^{2}\sigma\mathcal{V}_{YM}(\sigma)=\frac{1}{2\pi i}\oint\frac{\left\langle \lambda\xi\right\rangle }{\left\langle \lambda s\right\rangle \left\langle s\xi\right\rangle }\exp\left(\frac{\left\langle s\xi\right\rangle \left[s\mu\right]}{\left\langle \lambda\xi\right\rangle }\right)j,
\]
where the worldsheet coordinates being integrated over are $\sigma_{\alpha}=t^{-1}\left(1,\sigma\right)$. For an $(n+1$)-point N$^{k-2}$MHV amplitude, the contour
is localized around the $k-1$ solutions of $\left\langle s\lambda\left(\sigma\right)\right\rangle =0$,
and the positions of the hard vertex operators are determined
by the scattering equations for an $n$-point N$^{k-2}$MHV amplitude
as the momentum of the soft particle goes to zero, which is the approximation we use. One then wraps this contour around the location of the hard vertex operators. If we took a negative
helicity particle to be soft, the contour would be initially located around
the $n-k-1$ solutions of $\left[s\tilde{\lambda}\left(\sigma\right)\right]=0$,
and the positions of the hard vertex operators would be determined
by the scattering equations for an $n$-point N$^{k-3}$MHV amplitude
as the momentum of the soft particle goes to zero. For more details,
see Appendix \ref{4dscatt}. 

\begin{figure}[htbp] 
\centering
       \includegraphics[width=1.8in]{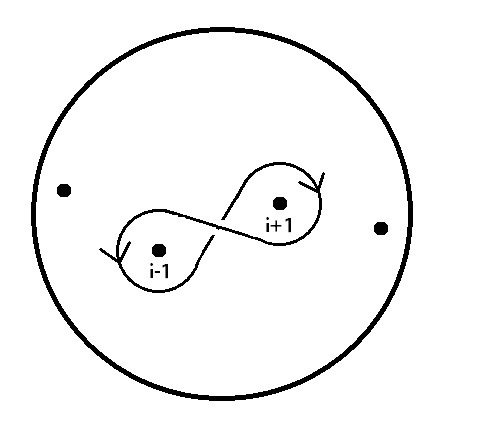}  
    \caption{If particle $i$ goes soft in a color-ordered Yang-Mills amplitude, the soft theorem follows from integrating its vertex operator around the vertex operators for particles $i-1$ and $i+1$, as depicted above for a genus zero worldsheet.}
    \label{ymb}
    \end{figure}

For color-ordered Yang-Mills amplitudes, the soft theorems follow from integrating the soft gluon vertex operator around the two adjacent hard vertex operators following a ``figure-eight'' path, as depicted
in Figure \ref{ymb}.
Let us describe this in more detail. Suppose that the soft
gluon has positive helicity and is adjacent to another positive helicity
gluon whose vertex operator is located at position $\sigma_i$. One then obtains the following
residue from integrating the soft gluon vertex operator around $\sigma_i$:
\begin{eqnarray} 
\int \rd^{2}\sigma\mathcal{V}_{YM}(\sigma)\mathcal{V}\left(\sigma_{i}\right)&=&\frac{1}{2\pi i}\oint\frac{\left\langle \lambda(\sigma)\xi\right\rangle }{\left\langle \lambda(\sigma)s\right\rangle \left\langle s\xi\right\rangle }\exp\left(\frac{\left\langle s\xi\right\rangle \left[s\mu(\sigma)\right]}{\left\langle \lambda(\sigma)\xi\right\rangle }\right)j(\sigma)\mathcal{V}\left(\sigma_{i}\right)
\nonumber \\
&= &
\frac{1}{2\pi i}\oint\frac{1}{\sigma-\sigma_{i}}\frac{\left\langle i\xi\right\rangle }{\left\langle is\right\rangle \left\langle s\xi\right\rangle }\exp\left(\frac{\left\langle s\xi\right\rangle }{\left\langle \lambda\xi\right\rangle }\tilde{\lambda}_{s}\cdot\frac{\partial}{\partial\tilde{\lambda}_{i}}\right)\mathcal{V}\left(\sigma_{i}\right)
\nonumber \\
&= &
\frac{\left\langle i\xi\right\rangle }{\left\langle is\right\rangle \left\langle s\xi\right\rangle }\exp\left(\frac{\left\langle s\xi\right\rangle }{\left\langle \lambda\xi\right\rangle }\tilde{\lambda}_{s}\cdot\frac{\partial}{\partial\tilde{\lambda}_{i}}\right)\mathcal{V}\left(\sigma_{i}\right).
\end{eqnarray}
In obtaining the second line, we kept the single
trace term in OPE of currents in \eqref{eq:ca} and noted that
\begin{equation}
\lim_{\sigma\rightarrow\sigma_{i}}\tilde{\lambda}(\sigma)\rightarrow\frac{t_{i}\tilde{\lambda}_{i}}{\sigma-\sigma_{i}},\,\,\,\lim_{\sigma\rightarrow\sigma_{i}}\lambda(\sigma)\rightarrow\frac{\lambda_{i}}{t_{i}},\,\,\,\lim_{\sigma\rightarrow\sigma_{i}}\mu(\sigma)\rightarrow\frac{\sigma-\sigma_{i}}{t_{i}}\frac{\partial}{\partial\tilde{\lambda}_{i}},\label{eq:limit}
\end{equation}
which follow from \eqref{eq:ope}, \eqref{scatts1}, and \eqref{scatts2}. 

Hence, if particle $n$ of an $n$-point color-ordered Yang-Mills amplitude has positive helicity, we find that 
\begin{multline}
\lim_{p_{n}^{+}\rightarrow0}\mathcal{A}_{n}^{YM}=\left[\frac{\left\langle 1\xi\right\rangle }{\left\langle 1n\right\rangle \left\langle n\xi\right\rangle }\exp\left(\frac{\left\langle n\xi\right\rangle }{\left\langle 1\xi\right\rangle }\tilde{\lambda}_{n}\cdot\frac{\partial}{\partial\tilde{\lambda}_{1}}\right)-\right.\\
\left.\frac{\left\langle n-1\xi\right\rangle }{\left\langle n-1 n\right\rangle \left\langle n\xi\right\rangle }\exp\left(\frac{\left\langle n\xi\right\rangle }{\left\langle \xi n-1\right\rangle }\tilde{\lambda}_{n}\cdot\frac{\partial}{\partial\tilde{\lambda}_{n-1}}\right)\right]\mathcal{A}_{n-1}^{YM}. \label{eq:softyma}
\end{multline}
If the soft particle particle has negative helicity, complex conjugate the prefactor.

\subsection{Gravity}

\begin{figure}[htbp] 
\centering
       \includegraphics[width=1.8in]{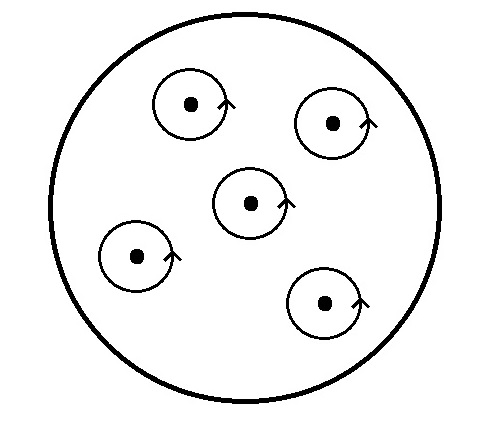}  
    \caption{The soft graviton theorem follows from integrating a soft graviton vertex operator around each of the hard vertex operators, as depicted above for a genus zero worldsheet.}
    \label{grb}
    \end{figure}

Now we will derive the soft graviton theorems. Similar manipulations to those described in the previous subsection imply that a positive helicity graviton vertex operator can be expressed as 
\[
\int \rd^{2}\sigma\mathcal{V}_{GR}(\sigma)=\frac{1}{2\pi i}\oint\frac{\left\langle \lambda\xi\right\rangle }{\left\langle \lambda s\right\rangle \left\langle s\xi\right\rangle }\left(\frac{\left[\tilde{\lambda}s\right]\left\langle \lambda\xi\right\rangle }{\left\langle s\xi\right\rangle }+i\left[\tilde{\rho}s\right]\left[\rho s\right]\right)\exp\left(\frac{\left\langle s\xi\right\rangle \left[s\mu\right]}{\left\langle \lambda\xi\right\rangle }\right).
\]
The soft graviton theorems then follow from integrating the soft graviton vertex operator around the location of each hard vertex operator in a correlation function and adding up the residues, as depicted in Figure \ref{grb}. 
The residue from integrating it around a hard vertex operator at $\sigma_i$ is given by
\[
\int \rd^{2}\sigma\mathcal{V}_{GR}(\sigma)\mathcal{V}\left(\sigma_{i}\right)=\frac{\left[is\right]\left\langle i\xi\right\rangle ^{2}}{\left\langle is\right\rangle \left\langle s\xi\right\rangle ^{2}}\exp\left(\frac{\left\langle s\xi\right\rangle }{\left\langle i\xi\right\rangle }\tilde{\lambda}_{s}\cdot\frac{\partial}{\partial\tilde{\lambda}_{i}}\right)\mathcal{V}\left(\sigma_{i}\right)
\]
where we used \eqref{eq:ope2} and \eqref{eq:limit}, and noted
that terms involving derivatives with respect to the worldsheet
fermions do not contribute because they contain a double pole. 

Hence, if particle $n$ of an $n$-point amplitude is a positive helicity graviton, we find
\begin{equation}
\lim_{p_{n}^{+}\rightarrow0}\mathcal{A}_{n}^{GR}=\sum_{i=1}^{n-1}\frac{\left[in\right]\left\langle i\xi\right\rangle ^{2}}{\left\langle in\right\rangle \left\langle n\xi\right\rangle ^{2}}\exp\left(\frac{\left\langle n\xi\right\rangle }{\left\langle i\xi\right\rangle }\tilde{\lambda}_{n}\cdot\frac{\partial}{\partial\tilde{\lambda}_{i}}\right)\mathcal{A}_{n-1}^{GR}\label{eq:softgra}
\end{equation}
If the soft particle has negative helicity, complex
conjugate the prefactor. Note that our derivation of this result did not make use of the detailed structure of the hard vertex operators and therefore reflects the universality of the soft graviton theorems. 

\subsection{Comparison to BCFW}

It is interesting to compare \eqref{eq:softyma} and \eqref{eq:softgra} to the results of BCFW recursion \cite{Cachazo:2014fwa,Casali:2014xpa,He:2014bga}:
\begin{equation}
\lim_{p_{n}^{+}\rightarrow0}\mathcal{A}_{n}^{YM}=\frac{\left\langle n-11\right\rangle }{\left\langle 1n\right\rangle \left\langle nn-1\right\rangle }\exp\left(\frac{\left\langle nn-1\right\rangle }{\left\langle 1n-1\right\rangle }\tilde{\lambda}_{n}\cdot\frac{\partial}{\partial\tilde{\lambda}_{1}}+\frac{\left\langle 1n\right\rangle }{\left\langle 1n-1\right\rangle }\tilde{\lambda}_{n}\cdot\frac{\partial}{\partial\tilde{\lambda}_{n-1}}\right)\mathcal{A}_{n-1}^{YM}\label{eq:softymbcfw}
\end{equation}
\begin{equation}
\lim_{p_{n}^{+}\rightarrow0}\mathcal{A}_{n}^{GR}=\sum_{i=1}^{n}\frac{\left[in\right]\left\langle in-1\right\rangle ^{2}}{\left\langle in\right\rangle \left\langle nn-1\right\rangle ^{2}}\exp\left(\frac{\left\langle nn-1\right\rangle }{\left\langle in-1\right\rangle }\tilde{\lambda}_{n}\cdot\frac{\partial}{\partial\tilde{\lambda}_{i}}+\frac{\left\langle ni\right\rangle }{\left\langle n-1i\right\rangle }\tilde{\lambda}_{n}\cdot\frac{\partial}{\partial\tilde{\lambda}_{n-1}}\right)\mathcal{A}_{n-1}^{GR}.\label{eq:softgrbcfw}
\end{equation}
Note that for non-MHV amplitudes, these formulae are only valid up to subleading
order in Yang-Mills and sub-subleading order in gravity. Let us first compare the the two soft graviton formulae. Choosing $\xi=\lambda_{n}$
in \eqref{eq:softgra} brings it into a very similar form to \eqref{eq:softgrbcfw}.
In fact, the leading terms trivially agree, and one can show that
the subleading and subsubleading terms also agree after using momentum
conservation \cite{Cachazo:2014fwa,He:2014bga}. Beyond subsubleading order however, \eqref{eq:softgra} and \eqref{eq:softgrbcfw} do not generally
agree. 

Now let us look at the soft gluon formulae. Expanding \eqref{eq:softyma}
to leading order and using the Schouten identity gives 
\[
\frac{\left\langle 1\xi\right\rangle }{\left\langle 1n\right\rangle \left\langle n\xi\right\rangle }-\frac{\left\langle n-1\xi\right\rangle }{\left\langle n-1n\right\rangle \left\langle n\xi\right\rangle }=\frac{\left\langle n-11\right\rangle }{\left\langle 1n\right\rangle \left\langle nn-1\right\rangle },
\]
which matches the leading order of \eqref{eq:softymbcfw}. Similarly,
expanding \eqref{eq:softyma} to subleading order, the dependence
on $\xi$ cancels out and one obtains 
\[
\frac{1}{\left\langle n1\right\rangle }\tilde{\lambda}_{n}\cdot\frac{\partial}{\partial\tilde{\lambda}_{1}}-\frac{1}{\left\langle nn-1\right\rangle }\tilde{\lambda}_{n}\cdot\frac{\partial}{\partial\tilde{\lambda}_{n-1}},
\]
which matches the subleading prefactor in \eqref{eq:softymbcfw}.
Beyond subleading order however, the two soft gluon formulae do not generally
agree. It was observed in \cite{He:2014bga} that the BCFW formulae should hold to all orders for MHV amplitudes. Hence we do not expect
\eqref{eq:softyma} and \eqref{eq:softgra} to hold beyond subleading
order for Yang-Mills amplitudes and sub-subleading order for
gravity amplitudes. 

In deriving \eqref{eq:softyma} and \eqref{eq:softgra}, we assumed that
the locations of the hard vertex operators did not depend on the location
of the soft vertex operator, i.e. they are solutions to $(n-1)$-point
scattering equations rather than $n$-point scattering equations.
It would be interesting to compute corrections to these formulae by
taking into account the ``backreaction'' of the soft vertex operator
on the the hard vertex operators via the scattering equations. 

In summary, we have proved the following soft theorems: 
\begin{itemize}
\item
For gravitational amplitudes,
\[
\lim_{p_{n}^{+}\rightarrow0}\mathcal{A}_{n}=\sum_{i=1}^{n-1}\left(S_{GR}^{(-1)}+S_{GR}^{(0)}+S_{GR}^{(1)}\right)\mathcal{A}_{n-1}
\]
\begin{equation}
S_{GR}^{(-1)}=\sum_{i=1}^{n-1}\frac{\left[in\right]\left\langle \xi i\right\rangle ^{2}}{\left\langle in\right\rangle \left\langle \xi n\right\rangle ^{2}},\,\,\, S_{GR}^{(0)}=\sum_{i=1}^{n}\frac{\left[in\right]\left\langle \xi i\right\rangle }{\left\langle in\right\rangle \left\langle \xi n\right\rangle }\tilde{\lambda}_{n}\cdot\frac{\partial}{\partial\tilde{\lambda}_{i}},\,\,\, S_{GR}^{(1)}=\frac{1}{2}\sum\frac{\left[in\right]}{\left\langle in\right\rangle }\tilde{\lambda}_{n}^{\dot{\alpha}}\tilde{\lambda}_{n}^{\dot{\beta}}\frac{\partial^{2}}{\partial\tilde{\lambda}_{i}^{\dot{\alpha}}\partial\tilde{\lambda}_{i}^{\dot{\beta}}}.
\label{grcheck}
\end{equation}
\item
For color-ordered YM amplitudes
\[
\lim_{p_{n}^{+}\rightarrow0}\mathcal{A}_{n}=\sum_{i=1}^{n-1}\left(S_{YM}^{(-1)}+S_{YM}^{(0)}\right)\mathcal{A}_{n-1}
\] 
\begin{equation}
S_{YM}^{(0)}=\frac{\left\langle n-11\right\rangle }{\left\langle n-1n\right\rangle \left\langle n1\right\rangle },\,\,\, S_{YM}^{(0)}=\frac{1}{\left\langle n1\right\rangle }\tilde{\lambda}_{n}\cdot\frac{\partial}{\partial\tilde{\lambda}_{1}}+\frac{1}{\left\langle n-1n\right\rangle }\tilde{\lambda}_{n}\cdot\frac{\partial}{\partial\tilde{\lambda}_{n}}.
\end{equation}
\item 
For photons, there is no color ordering so the soft theorems can be obtained by Taylor expanding the first term on the right hand side of \eqref{eq:softyma} to subleading order and summing over all of the hard legs:
\[
\lim_{p_{n}^{+}\rightarrow0}\mathcal{A}_{n}=\sum_{i=1}^{n-1}\left(S_{QED}^{(-1)}+S_{QED}^{(0)}\right)\mathcal{A}_{n-1}
\]
\begin{equation}
S_{QED}^{(-1)}=\sum_{i=1}^{n-1}\frac{\left\langle i\xi\right\rangle }{\left\langle in\right\rangle \left\langle n\xi\right\rangle },\,\,\, S_{QED}^{(0)}=\sum_{i=1}^{n-1}\frac{1}{\left\langle in\right\rangle }\tilde{\lambda}_{n}\cdot\frac{\partial}{\partial\tilde{\lambda}_{i}}.
\label{qes}
\end{equation}
\end{itemize}

\section{Symmetries and Braiding} \label{alsy} 

In this section, we explain how the soft theorems can be interpreted as Ward identities of ambitwistor string theory, and point out a simple relation betwen the charges which generate soft gluon theorems and those which generate soft graviton theorems. We then we explain how the algebra of soft limits can be encoded in the braiding of soft vertex operators on the worldsheet and how this reflects the underlying symmetry algebra of the scattering amplitudes. 

\subsection{Gravity vs Yang-Mills}
Taylor expanding the soft gluon and soft graviton vertex operators in \eqref{eq:ymvo} and \eqref{gv} in powers of the soft momentum gives: 
\[
\int \rd^{2}\sigma\mathcal{V}_{YM}(\sigma)=\sum_{l=-1}^{\infty}\frac{1}{(l+1)!}q_{YM}^{(l)}
\]
\[
\int \rd^{2}\sigma\mathcal{V}_{GR}(\sigma)=\sum_{l=-1}^{\infty}\frac{1}{(l+1)!}q_{GR}^{(l)}+\sum_{l=0}^{\infty}\frac{1}{l!}q_{\rho\tilde{\rho}}^{(l)}
\]
where

\begin{equation}
q_{YM}^{(l)}=\frac{1}{2\pi i}\oint\frac{1}{\left\langle s\lambda\right\rangle }\left(\frac{\left\langle \xi s\right\rangle }{\left\langle \xi\lambda\right\rangle }\right)^{l}\left[\mu s\right]^{l+1}j\label{eq:qym}
\end{equation}

\begin{equation}
q_{GR}^{(l)}=\frac{1}{2\pi i}\oint\frac{1}{\left\langle s\lambda\right\rangle }\left(\frac{\left\langle \xi s\right\rangle }{\left\langle \xi\lambda\right\rangle }\right)^{l-1}\left[\tilde{\lambda}s\right]\left[\mu s\right]^{l+1}\label{eq:qgr}
\end{equation}

\begin{equation}
q_{\rho\tilde{\rho}}^{(l)}=\frac{1}{2\pi i}\oint\frac{1}{\left\langle s\lambda\right\rangle }\left(\frac{\left\langle \xi s\right\rangle }{\left\langle \xi\lambda\right\rangle }\right)^{l-1}\left[\mu s\right]^{l}\left[\tilde{\rho}\tilde{\lambda}_{i}\right]\left[\rho\tilde{\lambda}_{i}\right].
\end{equation}
Each term in the Taylor expansion can be thought of as a charge on the worldsheet which generates a soft theorem when inserted into a correlation function. Indeed, the results of Section \ref{tree} imply that $\left\{ q_{YM}^{(-1)},q_{YM}^{(0)}\right\} $ generate the leading and subleading soft gluon theorems, while $\left\{ q_{GR}^{(-1)},q_{GR}^{(0)},q_{GR}^{(1)}\right\}$ generate the leading, subleading, and sub-subeading soft graviton theorems. Hence, the soft theorems can be interpreted as Ward identities of the 2d CFT describing ambitwistor string theory. This was also discussed in \cite{Geyer:2014lca}.  

Interestingly, the charges which generate soft gluon theorems are related in a simple way to those which generate soft graviton theorems. In particular, $q_{YM}^{(l)}$ can be mapped into $q_{GR}^{(l+1)}$ by replacing 
\[
j\rightarrow\left[\tilde{\lambda}s\right]\left[\mu s\right].
\]
In terms of OPE's this corresponds to replacing
\[
if_{\,\,\,\,\,\,\,\,\, C}^{AB}j^{C}\rightarrow\tilde{\lambda}_{s}^{\dot{\alpha}}\tilde{\lambda}_{s}^{\dot{\beta}}\tilde{m}_{\dot{\alpha}\dot{\beta}},\,\,\,\tilde{m}_{\dot{\alpha}\dot{\beta}}=\tilde{\lambda}_{(\dot{\alpha}}\frac{\partial}{\partial\tilde{\lambda}^{\dot{\beta})}}.
\]
Furthermore, $q_{YM}^{(l)}$ can be mapped into $q_{\rho\tilde{\rho}}^{(l+1)}$ by replacing
\[
j\rightarrow\left[\tilde{\rho}s\right]\left[\rho s\right].
\]  
Hence, we find that the charges which generate soft gluon theorems map into the charges which generate soft graviton theorems
after replacing the Kac-Moody current $j$ with a Lorentz generator. Note that this relates the leading
and subleading soft gluon theorems to the subleading and subsubleading
soft graviton theorems, which is natural since
the leading soft graviton theorem does not receive loop corrections,
but the subleading and subsubleading soft graviton theorems receive one and
two-loop corrections, respectively, just like the leading and subleading
soft gluon theorems. These results suggest a possible connection
to color-kinematics duality \cite{Bern:2008qj,Monteiro:2011pc}, which would be interesting to further explore.

\subsection{Soft Limit Algebra}

In this section, we will show that the algebra of soft limits can be elegantly encoded in the
braiding of soft vertex operators. To start off, consider a tree-level $n$-point
amplitude where particles $(n-1,n)$ are positive helicity gravitons
and consider taking particle $n-1$ soft followed by taking particle
$n$ soft. Using \eqref{grcheck} and keeping the leading order terms in the soft limits gives 

\begin{equation}
\sum_{i=1}^{n-2}\frac{\left[in\right]\left\langle \xi_{n}i\right\rangle ^{2}}{\left\langle in\right\rangle \left\langle \xi_{n}n\right\rangle ^{2}}\left(\sum_{j=1}^{n-2}\frac{\left[jn-1\right]\left\langle \xi_{n-1}j\right\rangle ^{2}}{\left\langle jn-1\right\rangle \left\langle \xi_{n-1}n-1\right\rangle ^{2}}+\frac{\left[nn-1\right]\left\langle \xi_{n-1}n\right\rangle ^{2}}{\left\langle nn-1\right\rangle \left\langle \xi_{n-1}n-1\right\rangle ^{2}}\right)\mathcal{A}_{n-2}.\label{eq:c1}
\end{equation}
On the other hand, if we first take particle $n$ soft and then take
particle $n-1$ soft we obtain 

\begin{equation}
\sum_{j=1}^{n-2}\frac{\left[jn-1\right]\left\langle \xi_{n-1}j\right\rangle ^{2}}{\left\langle jn-1\right\rangle \left\langle \xi_{n-1}n-1\right\rangle ^{2}}\left(\sum_{i=1}^{n-2}\frac{\left[in\right]\left\langle \xi_{n}i\right\rangle ^{2}}{\left\langle in\right\rangle \left\langle \xi_{n}n\right\rangle ^{2}}+\frac{\left[n-1n\right]\left\langle \xi_{n}n-1\right\rangle ^{2}}{\left\langle n-1n\right\rangle \left\langle \xi_{n}n\right\rangle ^{2}}\right)\mathcal{A}_{n-2}.\label{eq:c2}
\end{equation}
Subtracting \eqref{eq:c2} from \eqref{eq:c1} gives
\begin{multline}
\frac{\left[n-1n\right]}{\left\langle n-1n\right\rangle \left\langle \xi_{n-1}n-1\right\rangle ^{2}\left\langle \xi_{n}n\right\rangle ^{2}}
\\
\sum_{i=1}^{n-2}\left(\frac{\left\langle \xi_{n-1}n\right\rangle ^{2}\left[in\right]\left\langle \xi_{n}i\right\rangle ^{2}}{\left\langle in\right\rangle }-\frac{\left\langle \xi_{n}n-1\right\rangle ^{2}\left[in-1\right]\left\langle \xi_{n-1}i\right\rangle ^{2}}{\left\langle in-1\right\rangle }\right)\mathcal{A}_{n-2}.\label{eq:c3}
\end{multline}
This result can be easily understood from the perspective of CFT by recalling that the soft graviton theorems arise from integrating a soft graviton vertex operator around each of the
hard vertex operators. We then see that that the first term
in the parenthesis in \eqref{eq:c1} comes from braiding the vertex operator for particle
$n-1$ around the vertex operators for particles $1,...,n-2$, and the second term in the parenthesis
comes from braiding the vertex operator for particle $n-1$ around the vertex operator for particle $n$ before particle $n$ becomes soft. The sum to the left of the parenthesis then corresponds to braiding the vertex operator for particle
$n$ around the remaining $n-2$ hard vertex operators. There
is a similar interpretation for \eqref{eq:c2} if one exchanges the
roles of particles $n$ and $n-1$. 

\begin{figure}[htbp] 
\centering
       \includegraphics[width=3.5in]{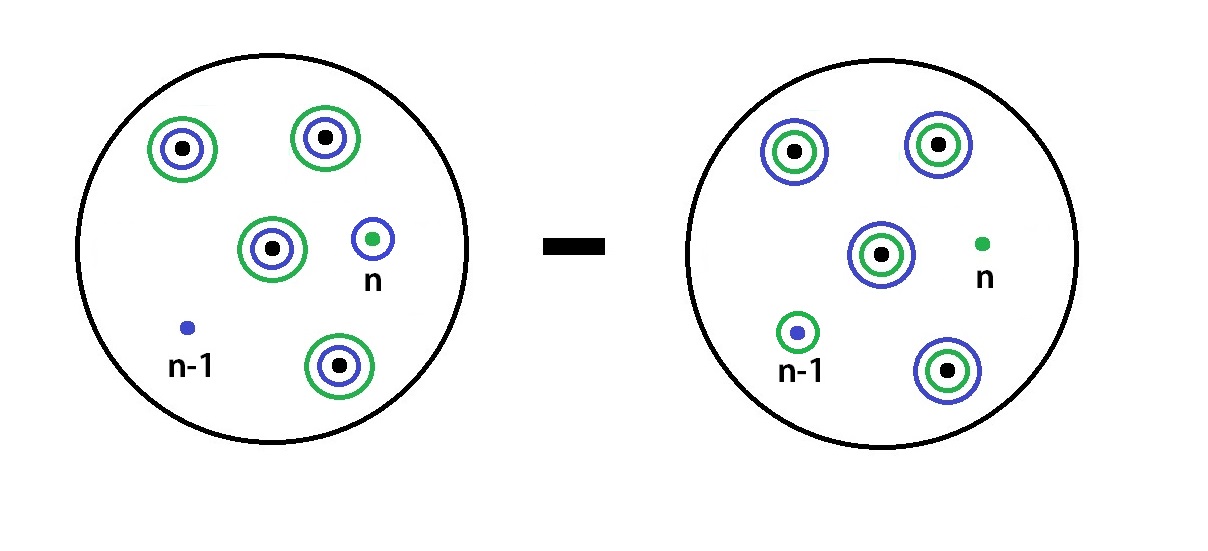}  
    \caption{The commutator of two soft graviton limits from the point of view of CFT. In the left diagram, particle $n-1$ goes soft before particle $n$, and in the right diagram, particle $n$ goes soft before particle $n-1$. The bulk contributions correspond to points with two circles around them and the boundary contributions correspond to points with one circle around them.}
    \label{gr2}
    \end{figure}

Hence, as depicted in Figure \ref{gr2},
there are two types of contributions to the commutator of soft
limits: the ``bulk'' contributions where both soft vertex operators braid
a hard vertex operator, and the ``boundary'' contributions where one soft vertex operator braids the other one before it becomes soft. In the example we are consdering, it is the second type of contribution which gives rise to \eqref{eq:c3}, and our prescription will be to discard such boundary terms. This can be achieved simply by choosing the reference spinors $(\xi_{n-1},\xi_n)=(\lambda_n,\lambda_{n-1})$, after which the commutator of soft limits vanishes. In the present example, the vanishing commutator can be understood as a consequence of Bose symmetry, since we are taking the two soft gravitons to have the same helicity, however using a similar calculation one finds that the commutator of soft limits also vanishes to leading order for soft gravitons of opposite helicity. This is no longer the case at subleading order, however \cite{Klose:2015xoa}. In Appendix \ref{details}, we show that the commutator of soft limits is nonzero at subleading order for photons and gravitons of opposite helicity.  

\begin{figure}[htbp] 
\centering
       \includegraphics[width=3in]{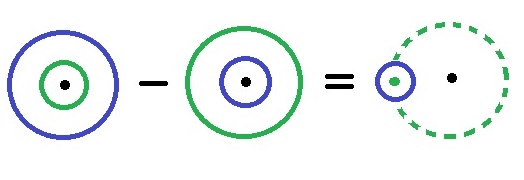}  
    \caption{The commutator of two soft limits can be encoded by braiding one soft vertex operator around another.}
    \label{br1}
    \end{figure}

Using the prescription defined above, the algebra of soft limits can be encoded in the braiding of soft vertex operators. This is essentially a consequence of the Jacobi identity: 
\begin{equation}
\left[q_{n},\left[q_{n-1},\mathcal{V}_{i}\right]\right]-\left[q_{n-1},\left[q_{n},\mathcal{V}_{i}\right]\right]=\left[\left[q_{n},q_{n-1}\right],\mathcal{V}_{i}\right],\label{eq:jac}
\end{equation}
where $(q_n,q_{n-1})$ are charges which generate soft limits for particles $(n,n-1)$ and $\mathcal{V}_i$ is a hard vertex operator. Indeed, recalling that time ordered commutators in a 2d CFT are described by contour integrals, we see that the left-hand-side of \eqref{eq:jac} corresponds to a bulk contribution to the commutator of the soft limits generated by $q_n$ and $q_{n-1}$, and the right-hand-side correponds to the soft limit generated by the charge which arises from braiding $q_n$ and $q_{n-1}$, as depicted in Figure \ref{br1}.
 
In summary, after discarding boundary terms, the commutator of soft limits can be encoded in the braiding of soft vertex operators as follows:
\begin{equation}
\left[\lim_{p_{n}\rightarrow0},\lim_{p_{n-1}\rightarrow0}\right]\mathcal{A}_{n}|_{bulk}=\int\Pi_{i=1}^{n-2}\rd^{2}\sigma_{i}\left\langle \left[q_{n},q_{n-1}\right]\mathcal{V}_{1}...\mathcal{V}_{n-2}\right\rangle. 
\label{braidlimit}
\end{equation} 
Note that this equation is schematic in that we are not specifying the helicity of the vertex operators or implementing gauge-fixing of the worldsheet theory in order to make the presentation as simple as possible, but it is straightforward to incorporate these details. In Appendix \ref{details}, we compute the subleading contribution to the commutator of two soft graviton limits and two soft photon limits and verify that it agrees with \eqref{braidlimit}.

The algebra of soft limits reflects the underlying symmetry algebra of the amplitudes. Indeed, we have shown that the commutator of two soft graviton limits vanishes at leading order, which is expected since at leading order these limits are generated by the charges $q_{GR}^{(-1)}$ in \eqref{eq:qgr}, which are abelian generators corresponding to supertranslations. On the other hand, commutators of subleading soft graviton limits no longer vanish in general since the the corresponding charges $q_{GR}^{(0)}$ are nonabelian.  Identifying the world-sheet fields $\lambda^{\alpha}$ with homogeneous coordinates of the 2-sphere at null infinity, we see that the charges $q_{GR}^{(0)}$ depend on both $\lambda$ and $\tilde{\lambda}$ and therefore do not correspond to conformal transformations of the 2-sphere. Hence, the subleading soft graviton theorem appears to be associated with diffeomorphisms which are more general than extended BMS transformations \cite{Campiglia:2015yka}. Note that these diffeomorphisms satisfy 
\begin{equation}
\frac{\partial}{\partial\lambda}\cdot\delta\lambda+\frac{\partial}{\partial\tilde{\lambda}}\cdot\delta\bar{\lambda}=0.\label{eq:ap}
\end{equation}
In particular, using the OPE's in \eqref{eq:ope}, we see that the transformations induced by $q_{GR}^{(0)}$ are 
\[
\delta_{GR}^{(0)}\lambda^{\alpha}=\frac{\left[\xi\tilde{\lambda}\right]\left\langle s\lambda\right\rangle }{\left[s\tilde{\lambda}\right]\left[\xi s\right]}\lambda_{s}^{\alpha},\,\,\,\delta_{GR}^{(0)}\tilde{\lambda}^{\dot{\alpha}}=\frac{\left\langle \xi\lambda\right\rangle \left[s\tilde{\lambda}\right]}{\left\langle s\lambda\right\rangle \left\langle \xi s\right\rangle }\tilde{\lambda}_{s}^{\dot{\alpha}}
\]
Hence, $\frac{\partial}{\partial\lambda}\cdot\delta_{GR}^{(0)}\lambda=\frac{\partial}{\partial\tilde{\lambda}}\cdot\delta_{GR}^{(0)}\tilde{\lambda}=0$.
Moreover, it is not difficult to see that the infinite set of diffeomorphisms generated commuting by these
transformations satisfy \eqref{eq:ap} as well.

\section{Loop Corrections} \label{loops}

In this section, we consider ambitwistor string theory on a genus one worldsheet. For this purpose, it is convenient
to work with the ambitwistor string theory developed by Mason and Skinner, whose target space can be defined in general dimensions \cite{Mason:2013sva}. First we review this model at tree-level, and then we describe the definition of genus one amplitudes in terms of the one-loop scattering equations developed in \cite{Adamo:2013tsa}. Finally, we use this formalism to compute the 1-loop correction to the subleading soft graviton theorem due to IR divergences.

\subsection{Ambitwsitor Strings in General Dimensions}
In this subsection we will review the ambitwistor string theory developed by Mason and Skinner, which can be used to describe gravity in any dimension. When $D=10$, this model becomes critical and its spectrum corresponds to type II supergravity (the existence of ambitwistor strings in general dimensions was also considered in \cite{Bandos:2014lja}). The model has $\mathcal{N}=2$  worldsheet supersymmetry and the Lagrangian is
\[
\mathcal{L}=p\cdot\bar{\partial}q+\sum_{r=1}^{2}\Psi_{r}\cdot\bar{\partial}\Psi_{r}+eT+\tilde{e}p^{2}+\sum_{r=1}^{2}\chi_{r}p\cdot\Psi_{r}
\]
where $p^{\mu}$ and $q_{\mu}$
are bosons with conformal weights $(1,0)$ and $(0,0)$, respectively, $\mu=0,...,D-1$ is a Lorentz index, 
$\Psi_{r}^{\mu}$ are fermions with conformal weight $\left(\frac{1}{2},0\right)$,
$(e,\tilde{e})$ are bosons with conformal weight $(-1,1)$, and $\chi_{r}^{\mu}$
are fermions with conformal weight $\left(-\frac{1}{2},1\right)$.

For a graviton with momentum $k^\mu$ and polarization $\epsilon^{\mu\nu}=\epsilon^{\mu}\epsilon^{\nu}$,
the integrated vertex operator is given by

\[
\mathcal{V}_{GR}=\delta\left(k\cdot p\right)e^{ik\cdot q}\Pi_{r=1}^{2}\left(\epsilon\cdot p+i\epsilon\cdot\Psi_{r}k\cdot\Psi_{r}\right).
\]
This is very similar to the definition of a graviton vertex operator in the RNS string
except for the delta function which ultimately gives rise to the
scattering equations. A tree-level $n$-point graviton
amplitude is computed from the correlation function

\[
\mathcal{A}_{n}=\int\frac{\Pi_{i=1}^{n}\rd^{2}\sigma_{i}}{\vol \, \SL(2,\C)}\left\langle \mathcal{V}_{1}...\mathcal{V}_{n}\right\rangle, 
\]
where the $\SL(2,\C)$ symmetry can be used to fix the location of three vertex operators.
Combining the exponentials with the action and integrating out the worldsheet
field $q^{\mu}$ gives 

\begin{equation}
p^\mu(\sigma)=\sum_{i=1}^{n}\frac{k^{\mu}_{i}}{\sigma-\sigma_{i}}.
\label{p0}
\end{equation}
Plugging this back into the delta functions implies the scattering
equations 

\[
\sum_{j\neq i}\frac{k_{i}\cdot k_{j}}{\sigma_{i}-\sigma_{j}}=0,\,\,\, i=1,...,n.
\]

The soft graviton theorems in general dimensions can be derived by Taylor expanding
a soft graviton vertex operator in the soft momentum \cite{Geyer:2014lca}. The
procedure is very similar to the one we described for the
4d ambitwistor string in Section \ref{tree}. Noting that 
\[
\delta\left(s\cdot p\right)=\frac{1}{2\pi i}\bar{\partial}\left(\frac{1}{s\cdot p}\right)
\]
and applying Stokes theorem, we obtain

\[
\int \rd^{2}\sigma\mathcal{V}_{GR}(\sigma)=\frac{1}{2\pi i}\oint\frac{e^{is\cdot q}}{s\cdot p}\Pi_{r=1}^{2}\left(\epsilon\cdot s+i\epsilon\cdot\Psi_{r}s\cdot\Psi_{r}\right),
\]
where $s$ is the momentum of the soft graviton. For an $n$-point amplitude, the contour encircles the $n-3$ solutions
of $s\cdot p\left(\sigma\right)=0$, and the positions of the hard
vertex operators are determined by the scattering equations of an $(n-1)$-point
amplitude as the momentum of the soft particle goes to zero, which is the approximation we use. We then wrap this contour around the locations of the hard vertex operators, as before. For more details about counting solutions,
see \cite{Cachazo:2013gna}. 

Expanding the soft graviton vertex operator in the soft momentum
gives an infinite series
\[
\int \rd^{2}\sigma\mathcal{V}_{GR}(\sigma)=\sum_{l=-1}^{\infty}\frac{1}{(l+1)!}q_{GR}^{(l)},
\]
where the leading term is
\[
q_{GR}^{(-1)}=\frac{1}{2\pi i}\oint\frac{(\epsilon\cdot p)^{2}}{s\cdot p}.
\]
Inserting this charge into correlation functions generates Weinberg's soft graviton
theorem. In particular, consider inserting this charge into an $n$-point
correlator. The soft theorem arises from integrating the charge around
each hard vertex operator and adding up the residues. The contribution
from the $i$th hard vertex operator is given by 

\begin{equation}
q_{GR}^{(-1)}\mathcal{V}\left(\sigma_{i}\right)=\frac{1}{2\pi i}\oint\frac{(\epsilon\cdot p(\sigma))^{2}}{s\cdot p(\sigma)}\mathcal{V}\left(\sigma_{i}\right)=\frac{\left(\epsilon\cdot k_{i}\right)^{2}}{s\cdot k_{i}}\mathcal{V}\left(\sigma_{i}\right)
\label{west}
\end{equation}
where we noted that $\lim_{\sigma\rightarrow\sigma_{i}}p^\mu(\sigma)=\frac{k^{\mu}_{i}}{\sigma-\sigma_{i}}$.

\subsection{One-Loop Scattering Equations}

In this section, we will review the proposal for computing 1-loop amplitudes using ambitwistor string theory, as described in \cite{Adamo:2013tsa,Casali:2014hfa}. When computing correlation functions on the
torus, the conformal isometries can be used to fix the location of one vertex operator. Suppose we fix the
location of the vertex operator for particle 1. In this case, the scattering equations for
an $n$-point amplitude are given by

\begin{equation}
k_{i}\cdot p(\sigma_{i})=0,\,\,\, i=2,3,...,n,\label{eq:scatteq1}
\end{equation}
\begin{equation}
p^{2}(\sigma_0)=0,\label{eq:scatteq2}
\end{equation}
where $k^{\mu}_i$ is the momentum of the $i$th particle, $\sigma_i$ is the location of the $i$th vertex operator, $\sigma_0$ is a point on the worldsheet which is in general distinct from the locations of the vertex operators, and $p^{\mu}(\sigma)$ is the generalization of \eqref{p0} to a toroidal worldsheet: 
\[
p^{\mu}(\sigma)=k^{\mu}+\sum_{i=1}^{n}k_{i}^{\mu}S\left(\sigma-\sigma_{i},\tau\right),\,\,\,S(\sigma,\tau)=\frac{\partial\theta_1(\sigma,\tau)}{\theta_{1}(\sigma,\tau)}+\frac{4\pi\Im \sigma}{\Im\tau},
\]
where $k^{\mu}$ corresponds to the loop momentum over which we integrate, and $\tau$ is the modular parameter of the torus.
Whereas \eqref{eq:scatteq1} determines the locations of the $n-1$ integrated vertex operators, \eqref{eq:scatteq2} determines $\tau$. More explicitly, the 1-loop scattering equations can be expressed as 

\begin{equation}
k\cdot k_{i}+\sum_{j\neq i}k_{i}\cdot k_{j}S\left(\sigma_{ij},\tau\right)=0,\,\,\, i=2,3,...,n,
\label{scat1}
\end{equation}
\begin{equation}
k^{2}+\sum_{i\neq j}k_{i}\cdot k_{j}f\left(\sigma_0,\sigma_{i},\sigma_{j},\tau\right)=0,\label{eq:p2}
\end{equation}
where $\sigma_{ij}=\sigma_i-\sigma_j$ and
\begin{equation}
f\left(\sigma_0,\sigma_{i},\sigma_{j},\tau\right)=S\left(\sigma_{0i},\tau\right)S\left(\sigma_{0j},\tau\right)-S\left(\sigma_{ij},\tau\right)S\left(\sigma_{0i},\tau\right)-S\left(\sigma_{ji},\tau\right)S\left(\sigma_{0j},\tau\right).
\label{f}
\end{equation}
One-loop amplitudes are then given by:
\begin{equation}
\mathcal{A}_{n}^{1-loop}=\int\frac{\rd^{D}k}{(2\pi)^{D-1}} \rd\tau\int\Pi_{i=2}^{n} \rd^{2}\sigma_{i} \delta\left(p^{2}(\sigma_0)\right) \left\langle \mathcal{V}_{1}...\mathcal{V}_{n}\right\rangle.  \label{eq:1loop}
\end{equation} 

\subsection{IR Divergences}

\begin{figure}[htbp] 
\centering
       \includegraphics[width=1.5in]{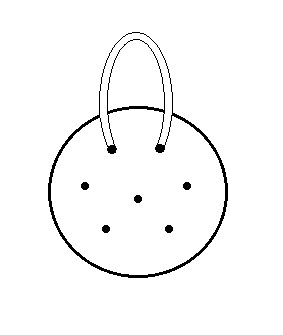}  
    \caption{A non-separating degeneration of a toroidal worldsheet gives rise to a spherical worldsheet with two additional punctures.}
    \label{pinch}
    \end{figure}

The IR divergent part of the genus one amplitude corresponds to the contribution
from $\Im\tau\rightarrow\infty$, or equivalently $q=e^{i\pi\tau}\rightarrow0$ \cite{Witten:2012bh}.
This corresponds to a non-separating degeneration of the toroidal worldsheet, which gives rise to a spherical world sheet with two additional
punctures, as shown in Figure \ref{pinch}. We will denote the location
of these punctures by $\sigma_{a}$ and $\sigma_{b}$. Using the conformal
symmetry of the 2-sphere, we can fix the locations of two more vertex operators, which we will take to correspond to particles 2 and 3 (note that the location of the vertex operator corresponding to particle 1 has already been fixed).

Equation \eqref{eq:p2} also simplifies in this limit. Noting that
\[
\frac{\partial\theta_1(\sigma,\tau)}{\theta_{1}(\sigma,\tau)}=\cot \sigma+4\sum_{n=1}^{\infty}\frac{q^{2n}}{1-q^{2n}}\sin(2n\sigma),
\]
one finds that the function $f$ in \eqref{f} reduces to $-1$ as $q\rightarrow 0$. In this limit, the delta function in \eqref{eq:1loop} therefore reduces to 
\[
\lim_{q\rightarrow0}\delta\left(p^{2}(\sigma_0)\right)=\delta\left(k^{2}-\sum_{i\neq j}k_{i}\cdot k_{j}\right)=\delta\left(k^{2}\right),
\]
where the sum runs over $\left\{ i,j\right\}  \in 1,...,n$, and we used momentum conservation to get the second equality. Hence, in this limit the loop momentum $k^{\mu}$ becomes on-shell, so we take the punctures at $\sigma_{a,b}$ to correspond to graviton vertex operators with external data $(\epsilon,k)$ and $(\epsilon^*,-k)$, respectively.  

The IR divergent part of the 1-loop amplitude corresponds to the soft limit $k^{\mu}\rightarrow0$. In this limit, we can expand the soft graviton vertex operators located at $\sigma_{a,b}$ in powers of the soft momentum to obtain   
\begin{equation}
\mathcal{A}_{n}^{1-loop}|_{div}=\int\frac{\rd^{D}k}{(2\pi)^{D-1}}\delta\left(k^{2}\right)\sum_{\pm}\Pi_{i=4}^{n} \rd^{2}\sigma_{i}\left\langle q_{a}^{(-1)}q_{b}^{(-1)}\mathcal{V}_{1}...\mathcal{V}_{n}\right\rangle _{tree}
\label{q0}
\end{equation}
where $\sum_{\pm}$ is the sum over soft graviton polarizations and
\[
q^{(-1)}_{a}=\frac{1}{2\pi i}\oint\frac{(\epsilon\cdot p)^{2}}{k\cdot p},\,\,\, q^{(-1)}_{b}=\frac{1}{2\pi i}\oint\frac{(\epsilon^{*}\cdot p)^{2}}{-k\cdot p}.
\]
Hence, the IR divergent part of the loop integrand arises from taking a double soft limit of a tree-level amplitude. Using \eqref{west} and the prescription for computing double soft limits described in Section \ref{alsy}, we obtain 
\[
\left\langle q^{(-1)}_{a} q^{(-1)}_{b}\mathcal{V}_{1}...\mathcal{V}_{n}\right\rangle _{tree}=\frac{1}{2}\sum_{i,j}\sum_{\pm}\frac{\left(\epsilon\cdot k_{i}\right)^{2}}{k\cdot k_{i}}\frac{\left(\epsilon^{*}\cdot k_{j}\right)^{2}}{-k\cdot k_{j}}\left\langle \mathcal{V}_{1}...\mathcal{V}_{n}\right\rangle _{tree},
\]
where the sum is over all pairs of vertex operators which are encircled
by $q^{(-1)}_{a,b}$, and the factor of $\frac{1}{2}$ comes from Bose symmetry. Hence, we find that
\begin{equation}
\left.\mathcal{A}_{n}^{1-loop}\right|_{div}=\left[\frac{1}{2}\int\frac{\rd^{D}k}{(2\pi)^{D-1}}\delta\left(k^{2}\right)\sum_{i,j}\sum_{\pm}\frac{\left(\epsilon\cdot k_{i}\right)^{2}}{k\cdot k_{i}}\frac{\left(\epsilon^{*}\cdot k_{j}\right)^{2}}{-k\cdot k_{j}}\right]\mathcal{A}_{n}^{tree}
\label{1-loopIR}
\end{equation}
This equation has a simple physical interpretation. It corresponds to the IR divergence which comes from a virtual graviton becoming on-shell. Weinberg showed that this divergence exponentiates and is cancelled by the real IR divergence coming from external soft gravitons \cite{Weinberg:1965nx}. From the point of view of ambitwistor string theory, the on-shell condition arises from the genus one scattering equations as $\Im \tau \rightarrow \infty$, and the sum over pairs of soft factors comes from braiding the pair of soft graviton vertex operators corresponding to the endpoints of the thin tube in Figure \ref{pinch} around the other vertex operators. 

Let us now evaluate the integral in equation \eqref{1-loopIR}. Noting that
\[
\sum_{\pm}\epsilon_{\mu}\epsilon_{\nu}\epsilon_{\rho}^{*}\epsilon_{\sigma}^{*}\rightarrow\frac{1}{2}\left(\eta_{\mu\rho}\eta_{\nu\sigma}+\eta_{\mu\sigma}\eta_{\nu\rho}-\eta_{\mu\nu}\eta_{\rho\sigma}\right)
\]
and using the identity $\delta\left(k^{2}\right)=-\frac{1}{\pi}\lim_{\varepsilon \rightarrow 0}\Im\left(\frac{1}{k^{2}+i\varepsilon}\right)$, \eqref{1-loopIR} reduces to
\[
\left.\mathcal{A}_{n}^{1-loop}\right|_{div}=\lim_{\varepsilon \rightarrow 0}\Im\left[\int\frac{\rd^{D}k}{(2\pi)^{D}}\sum_{i \neq j}\frac{\left(k_{i}\cdot k_{j}\right)^{2}}{k\cdot k_{i}k\cdot k_{j}\left(k^{2}+i\varepsilon\right)}\right]\mathcal{A}_{n}^{tree},
\] 
where we noted that the external momenta are null. As we mentioned above, this integral is IR divergent in four dimensions and needs to be regulated. Taking $D=4+2\epsilon$ with $\epsilon>0$ gives
\begin{equation}
\left.\mathcal{A}_{n}^{1-loop}\right|_{div}=\tilde{\sigma}_{n}\mathcal{A}_{n}^{tree},\label{eq:softm}
\end{equation}
where
\[
\tilde{\sigma}_{n}=-\frac{1}{16\pi^{2}\epsilon}\sum_{i\neq j}^{n}s_{ij}\ln\left(-\frac{\mu^{2}}{s_{ij}}\right),
\]
$s_{ij}=(p_i+p_j)^2$, and $\mu$ is an arbitrary energy scale \cite{Naculich:2011ry}. Note that double poles in $\epsilon$ which originate from collinear IR diveregences cancel out by momentum conservation.

Using \eqref{eq:softm}, it is not difficult to compute the IR divergent part of the 1-loop correction to the subleading soft graviton theorem, as shown in \cite{Bern:2014oka}. In particular, taking particle $n$ to be
soft and expanding \eqref{eq:softm} to subleading order in the soft momentum gives
\[
\tilde{\sigma}_{n}\rightarrow \tilde{\sigma}_{n-1}+\tilde{\sigma}_{n}',\,\,\,\mathcal{A}_{n}^{tree}\rightarrow\left(S^{(-1)}+S^{(0)}\right)\mathcal{A}_{n-1}^{tree},
\]
where
\[
\tilde{\sigma}_{n}'=-\frac{1}{16\pi^{2}\epsilon}\sum_{i=1}^{n-1}s_{in}\ln\left(-\frac{\mu^{2}}{s_{in}}\right)
\]
and $(S^{(-1)},S^{(0)})$ are given by  
\[
S^{(-1)}=\sum_{i=1}^{n-1}\frac{\left(\epsilon\cdot k_{i}\right)^{2}}{k_{n}\cdot k_{i}},\,\,\, S^{(0)}=\sum_{i=1}^{n-1}\frac{\epsilon\cdot k_{i}k_{n,\mu}\epsilon_{\nu}J_{i}^{\mu\nu}}{k_{n}\cdot k_{i}},
\]
where $J_i^{\mu\nu}=k_{i}^{[\mu}\frac{\partial}{\partial k_{i,\nu]}}+\epsilon_{i}^{[\mu}\frac{\partial}{\partial\epsilon_{i,\nu]}}$. Hence, when particle $n$ becomes soft, expanding \eqref{eq:softm} to subleading order in the soft momentum
gives
\[
\left.\mathcal{A}_{n}^{1-loop}\right|_{div}\rightarrow \tilde{\sigma}_{n-1}\left(S^{(-1)}+S^{(0)}\right)\mathcal{A}_{n-1}^{tree}+\tilde{\sigma}_{n}'S^{(-1)}\mathcal{A}_{n-1}^{tree}.
\]
This equation can alternatively be written as follows:
\[
\left.\mathcal{A}_{n}^{1-loop}\right|_{div}\rightarrow\left(S^{(-1)}+S^{(0)}\right)\left.\mathcal{A}_{n-1}^{1-loop}\right|_{div}+\left.S^{(0)1-loop}\right|_{div}\mathcal{A}_{n-1}^{tree},
\]
where
\[
\left.S^{(0)1-loop}\right|_{div}=\tilde{\sigma}_{n}'S^{(-1)}-\left(S^{(0)}\tilde{\sigma}_{n-1}\right).
\]
From this result, we see that the leading soft graviton theorem is not renormalized and the subleading soft graviton theorem
is renormalized at 1-loop, which is consistent with dimensional analysis.

\section{Conclusion}

In this paper, we used ambitwistor string theory to obtain formulae for the soft limits of scattering amplitudes of pure Yang-Mills theory and Einstein gravity in the form of an infinite series in the soft momentum. These formulae were derived by taking a vertex operator in a correlator to be soft, and evaluating the correlator in the limit that the scattering equations for the hard vertex operators are independent of the location of the soft vertex operator, which is valid up to subleading order in Yang-Mills theory and sub-subleading order in gravity. It would be interesting to compute higher order soft terms by taking into account the backreaction of the soft vertex operator on the hard vertex operators via the scattering equations, and to see if the results are constrained by symmetries such as conformal, Lorentz, and gauge invariance \cite{Larkoski:2014hta,Broedel:2014fsa,Bern:2014vva,Broedel:2014bza}. 

Each term in the Taylor expansion of a soft vertex operator can be thought of as a charge on the worldsheet which gives rise to Ward identities when insterted into correlation functions. Hence, the soft theorems can be interpreted as Ward identities for ambitwistor string theory, at least up to subleading order in Yang-Mills theory and subsubleading order in gravity. We have demonstrated that the algebra of soft limits can be encoded in the braiding of soft vertex operators if one discards boundary terms which arise from braiding one soft vertex operator around another one before it becomes soft. In this way, we find that the algebra associated with the leading order soft graviton theorem is abelian, furnishing a representation of supertranslations. On the other hand, the algebra associated with higher order soft graviton theorems is nonabelian and appears to be more general than the extended BMS algebra. There is also a similar story for Yang-Mills theory. In particular, the Kac-Moody symmetry of Yang-Mills amplitudes recently discussed in \cite{Strominger:2013lka,He:2015zea} is encoded by the Kac-Moody current $j$ which appears in the definition of the gluon vertex operators of ambitwistor string theory. Moreover, we find that the worldsheet charges which generate soft gluon theorems can be mapped into those which generate soft graviton theorems by replacing $j$ with a Lorentz generator, suggesting a possible connection to color-kinematics duality which would be be interesting to study further.

Finally, we computed the 1-loop IR divergent correction to the subleading soft graviton theorem by considering ambitwistor string theory on a genus one worldsheet, providing further evidence that the loop amplitudes of ambitwistor string theory correspond to field theory loop amplitudes. It would be interesting to demonstrate this beyond the IR limit, which is challenging because the 1-loop scattering equations and loop integrand of ambitwistor string theory generically contain elliptic functions. In ordinary string theory, modular invariance of one-loop amplitudes arises from summing over an infinite tower of states, and integrating over the real part of the modular parameter of the worldsheet implements level-matching, both of which should be absent in ambitwistor string theory. Hence, it would be desirable to extend the ambitwistor string framework to loop-level in such a way that makes these properties manifest.

It is remarkable that the soft theorems of pure Yang-Mills theory and Einstein gravity arise as Ward identities of a 2d CFT at null infinity. This suggests the possibility that these theories exhibit some form of integrability even though they do not possess Yangian symmetry. Given that ambitwistor string theory provides a concrete realization of the 2d CFT at null infinity, it should be a powerful tool for exploring this direction.

\begin{center}
\textbf{Acknowledgements}
\end{center}

This work was supported by the German Science Foundation (DFG) within the Collaborative Research Center 676 "Particles, Strings and the Early Universe". We thank Tim Adamo, Rutger Boels, Eduardo Casali, Wellington Galleas, Yvonne Geyer, Lionel Mason, Ricardo Monteiro, and Volker Schomerus for stimulating conversations.

\appendix 

\section{4d Scattering Equations} \label{4dscatt}

In this appendix, we will describe some basic properties of the tree-level
4d scattering equations. These equations are implicit in Witten's parity invariant formulation of twistor string theory \cite{Witten:2004cp}.
For an $n$-point amplitude where the first $k$ particles have negative
helicity and the last $n-k$ particles have positive helicity, the
scattering equations read

\begin{equation}
\sum_{i=1}^{k}\frac{t_{i}\left\langle pi\right\rangle }{\sigma_{p}-\sigma_{i}}=0,\,\,\, p\in\left\{ k+1,...,n\right\} ,\,\,\,\sum_{p=k+1}^{n}\frac{t_{p}\left[ip\right]}{\sigma_{i}-\sigma_{p}}=0,\,\,\, i\in\left\{ 1,...,k\right\} .\label{eq:4dse}
\end{equation}
Note that these equations imply the following conditions on the $t$
variables:
\begin{equation}
\sum_{i=1}^{k}t_{i}\left\langle pi\right\rangle =0,\,\,\,\sum_{p=k+1}^{n}t_{p}\left[ip\right]=0.\label{eq:constraints}
\end{equation}
To see this, perform a contour integral of the scattering equations
with the contour placed at infinity and add up all the residues. First
we will prove that the scattering equations are invariant under an
$\SL(2,\C)$ transformation of the worldsheet coordinates, which reflects the conformal symmetry of the underlying worldsheet theory:

\[
\sigma\rightarrow\frac{A\sigma+B}{C\sigma+D},\,\,\, AD-BC=1.
\]
Noting that under an $\SL(2,\C)$ transformation

\[
\sigma_{i}-\sigma_{j}\rightarrow\frac{\sigma_{i}-\sigma_j}{\left(C\sigma_{i}+D\right)\left(C\sigma_{j}+D\right)}
\]
we see that

\[
\sum_{i=1}^{k}\frac{t_{i}\left\langle pi\right\rangle }{\sigma_{p}-\sigma_{i}}\rightarrow\left(C\sigma_{p}+D\right)\sum_{i=1}^{k}\frac{t_{i}\left\langle pi\right\rangle \left(C\sigma_{i}+D\right)}{\sigma_{p}-\sigma_{i}}=\left(C\sigma_{p}+D\right)^{2}\sum_{i=1}^{k}\frac{t_{i}\left\langle pi\right\rangle }{\sigma_{p}-\sigma_{i}},
\]
which shows that the first set of equations in \eqref{eq:4dse} are
invariant. To obtain the equality above, we noted that

\[
\sum_{i=1}^{k}\frac{t_{i}\left\langle pi\right\rangle \sigma_{i}}{\sigma_{p}-\sigma_{i}}=-\sum_{i=1}^{k}t_{i}\left\langle pi\right\rangle +\sigma_{p}\sum_{i=1}^{k}\frac{t_{i}\left\langle pi\right\rangle }{\sigma_{p}-\sigma_{i}}
\]
and used \eqref{eq:constraints}. Using a similar calculation, one
finds that the second set of equations in \eqref{eq:4dse} are also
invariant under $\SL(2,\C)$ transformations. Using $\SL(2,\C)$ symmetry, one
can fix the position of three punctures. Hence, the scattering equations are
trivial for three-point amplitudes. For a four-particle MHV amplitude, they
reduce to a linear eqution for $\sigma_{4}$ after fixing $\left\{ \sigma_{1},\sigma_{2},\sigma_{3}\right\} =\left\{ 0,1,\infty\right\} $. Hence, there is only one solution
for $n=4$.

We will now show that for an $n$-point N$^{k-2}$MHV amplitude, there
are $A(n-3,k-2)$ solutions, where $A(i,j)$ are the Eulerian numbers
which are defined recursively as follows: 

\[
A(i,j)=(i-j)A(i-1,j-1)+(j+1)A(i-1,j),
\]
where $A(i,j)=0$ if $j<0$ or $j>i$, and $A(1,0)=1$. We will prove
this by induction. Using $\SL(2,\C)$ symmetry of the scattering equations,
we have already shown that when $n=4$ and $k=2$, the number of solutions
is $A(1,0)=1$. Now assume that the formula for the number of solutions
is true for all $(n-1)$-point amplitudes and consider an $n$-point
N$^{k-2}$MHV amplitude where the first $k$ particles have negative
helicity and the last $n-k$ particles have positive helicity, for
which the scattering equations are given by \eqref{eq:4dse}. Rescaling
the momenta of particles $k$ and $n$ according to $(p_{k},p_{n})\rightarrow\left(\tilde{\epsilon}p_{k},\epsilon p_{n}\right)$
and taking the limit $(\tilde{\epsilon},\epsilon)\rightarrow(0,1)$,
the scattering equations reduce to

\[
\sum_{i=1}^{k-1}\frac{t_{i}\left\langle pi\right\rangle }{\sigma_{p}-\sigma_{i}}=0,\,\,\, p\in\left\{ k+1,...,n\right\} ,\,\,\,\sum_{p=k+1}^{n}\frac{t_{j}\left[ip\right]}{\sigma_{i}-\sigma_{p}}=0,\,\,\, i\in\left\{ 1,...,k\right\} .
\]
From these equations, we see that all of the $\sigma_{i\neq k}$ are
determined by scattering equations for an $(n-1)$-point N$^{k-3}$MHV
amplitude, and for each solution, the equation for $\sigma_{k}$
has $n-k-1$ solutions. Now consider the limit $(\tilde{\epsilon},\epsilon)\rightarrow(1,0)$:

\[
\sum_{i=1}^{k}\frac{t_{i}\left\langle pi\right\rangle }{\sigma_{p}-\sigma_{i}}=0,\,\,\, p\in\left\{ k+1,...,n\right\} ,\,\,\,\sum_{p=k+1}^{n-1}\frac{t_{j}\left[ip\right]}{\sigma_{i}-\sigma_{p}}=0,\,\,\, i\in\left\{ 1,...,k\right\} 
\]
In this case, all of the $\sigma_{i\neq n}$ are determined by scattering
equations for an $(n-1)$-point N$^{k-2}$MHV amplitude, and for each solution, the equation for $\sigma_{n}$ has $k-1$ solutions.
Since the number of solutions will not change as $(\epsilon,\tilde{\epsilon})$
are varied smoothly, we see that when $\tilde{\epsilon}=\epsilon=1$
the number of solutions is $(n-k-1)A(n-4,k-3)+(k-1)A(n-4,k-2)=A(n-3,k-2)$,
where we have applied the inductive hypothesis and used the recursive
definition of Eulerian numbers. 

\section{Examples of Soft Limit Algebra} \label{details}

In Section \ref{alsy}, we described how the
algebra of soft limits can be encoded in the braiding of soft vertex
operators, if one discards terms which correspond to braiding one soft vertex operator around another one before it becomes soft, which we refer to as boundary terms. We refer to the remaining contributions as bulk terms. In this appendix, we will describe some nontrivial examples. In particular, we will use this prescription to compute the subleading contribution to the commutator of two soft graviton limits and two soft photon limits, and match these results with \eqref{braidlimit} (note that the leading contribution vanishes in both cases).

\subsection{Gravity}
Let us consider the commutator of two soft graviton limits, neglecting boundary terms. Since the commutator of two leading order soft limits vanishes regardless of the helicity of the soft gravitions, we will focus on terms which contain the leading order soft limit for one particle and the subleading order soft limit for the other. In this case, the commutator is nonzero if the two soft gravitons have opposite helicity. 

To start off, consider a tree-level $n$-point amplitude where particles $(n-1,n)$ are positive helicity gravitons and consider taking particle $n-1$
soft followed by taking particle $n$ soft:
\begin{multline}
\sum_{i=1}^{n-2}\frac{\left[in\right]\left\langle \xi_{n}i\right\rangle ^{2}}{\left\langle in\right\rangle \left\langle \xi_{n}n\right\rangle ^{2}}\sum_{j=1}^{n-2}\frac{\left[jn-1\right]\left\langle \xi_{n-1}j\right\rangle }{\left\langle jn-1\right\rangle \left\langle \xi_{n-1}n-1\right\rangle }\tilde{\lambda}_{n-1}\cdot\frac{\partial}{\partial\tilde{\lambda}_{j}}\mathcal{A}_{n-2}+
\\
\sum_{i=1}^{n-2}\frac{\left[in\right]\left\langle \xi_{n}i\right\rangle }{\left\langle in\right\rangle \left\langle \xi_{n}n\right\rangle }\tilde{\lambda}_{n}\cdot\frac{\partial}{\partial\tilde{\lambda}_{i}}\left(\sum_{j=1}^{n-2}\frac{\left[jn-1\right]\left\langle \xi_{n-1}j\right\rangle ^{2}}{\left\langle jn-1\right\rangle \left\langle \xi_{n-1}n-1\right\rangle ^{2}}\mathcal{A}_{n-2}\right),
\label{a1}
\end{multline}
where we used \eqref{grcheck}. Similarly, taking the soft limits in the reverse order gives
\begin{multline}
\sum_{j=1}^{n-2}\frac{\left[jn-1\right]\left\langle \xi_{n-1}j\right\rangle }{\left\langle jn-1\right\rangle \left\langle \xi_{n-1}n-1\right\rangle }\tilde{\lambda}_{n-1}\cdot\frac{\partial}{\partial\tilde{\lambda}_{j}}\left(\sum_{i=1}^{n-2}\frac{\left[in\right]\left\langle \xi_{n}i\right\rangle ^{2}}{\left\langle in\right\rangle \left\langle \xi_{n}n\right\rangle ^{2}}\mathcal{A}_{n-2}\right)+
\\
\sum_{j=1}^{n-2}\frac{\left[jn-1\right]\left\langle \xi_{n-1}j\right\rangle ^{2}}{\left\langle jn-1\right\rangle \left\langle \xi_{n-1}n-1\right\rangle ^{2}}\sum_{i=1}^{n-2}\frac{\left[in\right]\left\langle \xi_{n}i\right\rangle }{\left\langle in\right\rangle \left\langle \xi_{n}n\right\rangle }\tilde{\lambda}_{n}\cdot\frac{\partial}{\partial\tilde{\lambda}_{i}}\mathcal{A}_{n-2}.
\label{a2}
\end{multline}
Subtracting \eqref{a2} from \eqref{a1} then gives 
\begin{multline}
\left[\lim_{p_{n}^{+}\rightarrow0},\lim_{p_{n-1}^{+}\rightarrow0}\right]\mathcal{A}_{n}|_{bulk}=\frac{\left[nn-1\right]}{\left\langle \xi_{n}n\right\rangle \left\langle \xi_{n-1}n-1\right\rangle }\sum_{i=1}^{n-2}\left(\frac{\left[in-1\right]\left\langle \xi_{n-1}i\right\rangle \left\langle \xi_{n}i\right\rangle ^{2}}{\left\langle in\right\rangle \left\langle in-1\right\rangle \left\langle \xi_{n}n\right\rangle }+\right.
\\
\left.\frac{\left[in\right]\left\langle \xi_{n}i\right\rangle \left\langle \xi_{n-1}i\right\rangle ^{2}}{\left\langle in\right\rangle \left\langle in-1\right\rangle \left\langle \xi_{n-1}n-1\right\rangle }\right)\mathcal{A}_{n-2},
\label{eq:subc}
\end{multline}
where $|_{bulk}$ indicates that we are discarding boundary contributions to the commutator. Choosing the reference spinors $(\xi_{n-1},\xi_n)=(\lambda_n,\lambda_{n-1})$, we see that the commutator vanishes by momentum conservation.

Next, let's compute the commutator of soft limits in the case where the soft gravitons have opposite helicity, which was considered in \cite{Klose:2015xoa}.  In particular, suppose that particle $n$ has positive helicity and particle $n-1$ has negative helicity, and consider taking particle $n-1$ soft followed by taking particle $n$ soft:
\begin{multline}
\sum_{i=1}^{n-2}\frac{\left[in\right]\left\langle \xi_{n}i\right\rangle ^{2}}{\left\langle in\right\rangle \left\langle \xi_{n}n\right\rangle ^{2}}\sum_{j=1}^{n-2}\frac{\left\langle jn-1\right\rangle \left[\xi_{n-1}j\right]}{\left[jn-1\right]\left[\xi_{n-1}n-1\right]}\lambda_{n-1}\cdot\frac{\partial}{\partial\lambda_{j}}\mathcal{A}_{n-2}+
\\
\sum_{i=1}^{n-2}\frac{\left[in\right]\left\langle \xi_{n}i\right\rangle }{\left\langle in\right\rangle \left\langle \xi_{n}n\right\rangle }\tilde{\lambda}_{n}\cdot\frac{\partial}{\partial\tilde{\lambda}_{i}}\left(\sum_{j=1}^{n-2}\frac{\left\langle jn-1\right\rangle \left[\xi_{n-1}j\right]^{2}}{\left[jn-1\right]\left[\xi_{n-1}n-1\right]^{2}}\mathcal{A}_{n-2}\right).
\label{c1}
\end{multline}
Taking the soft limits in the reverse order gives
\begin{multline}
\sum_{j=1}^{n-2}\frac{\left\langle jn-1\right\rangle \left[\xi_{n-1}j\right]}{\left[jn-1\right]\left[\xi_{n-1}n-1\right]}\lambda_{n-1}\cdot\frac{\partial}{\partial\lambda_{j}}\left(\sum_{i=1}^{n-2}\frac{\left[in\right]\left\langle \xi_{n}i\right\rangle ^{2}}{\left\langle in\right\rangle \left\langle \xi_{n}n\right\rangle ^{2}}\mathcal{A}_{n-2}\right)+
\\
\sum_{j=1}^{n-2}\frac{\left\langle jn-1\right\rangle \left[\xi_{n-1}j\right]^{2}}{\left[jn-1\right]\left[\xi_{n-1}n-1\right]^{2}}\sum_{i=1}^{n-2}\frac{\left[in\right]\left\langle \xi_{n}i\right\rangle }{\left\langle in\right\rangle \left\langle \xi_{n}n\right\rangle }\tilde{\lambda}_{n}\cdot\frac{\partial}{\partial\tilde{\lambda}_{i}}\mathcal{A}_{n-2}.
\label{c2}
\end{multline}
Subtracting \eqref{c2} from \eqref{c1}, choosing the reference spinors $(\xi_{n-1},\xi_n)=(\lambda_n,\lambda_{n-1})$, and simplifying using the Schouten identity then gives 
\begin{equation}
\left[\lim_{p_{n}^{+}\rightarrow0},\lim_{p_{n-1}^{-}\rightarrow0}\right]\mathcal{A}_{n}|_{bulk}=\frac{1}{\left\langle n-1n\right\rangle \left[nn-1\right]}\sum_{i=1}^{n-2}\frac{\left[in\right]^{2}\left\langle in-1\right\rangle ^{2}}{\left\langle in\right\rangle ^{2}\left[in-1\right]^{2}}\left[i\right|p_{n-1}-p_{n}\left|i\right\rangle \mathcal{A}_{n-2},
\label{grpm}
\end{equation}
which agrees with the result obtained in \cite{Klose:2015xoa}.

Now we will derive equations \eqref{eq:subc} and \eqref{grpm} from the point of view of CFT by braiding two soft graviton vertex operators and inserting the resulting charge into an $(n-2)$-point correlation function of hard vertex operators. Since \eqref{eq:subc} describes the case where both soft gravitons have positive helicity, it arises from braiding the following soft graviton charges defined in \eqref{eq:qgr}:
\begin{equation}
\left[q_{n}^{(-1)},q_{n-1}^{(0)}\right]+\left[q_{n}^{(0)},q_{n-1}^{(-1)}\right],
\label{commutator}
\end{equation}
where
\[
q_{n}^{(-1)}=\frac{1}{2\pi i}\oint\frac{\left\langle \xi_{n}\lambda\right\rangle ^{2}\left[\tilde{\lambda}n\right]}{\left\langle n\lambda\right\rangle \left\langle \xi_{n}n\right\rangle ^{2}},\,\,\, q_{n}^{(0)}=\frac{1}{2\pi i}\oint\frac{\left\langle \xi_{n}\lambda\right\rangle \left[\tilde{\lambda}n\right]\left[\mu n\right]}{\left\langle \xi_{n}n\right\rangle \left\langle n\lambda\right\rangle }.
\]
Using the OPE's in \eqref{eq:ope}, one finds that \eqref{commutator}
is given by
\[
\frac{1}{2\pi i}\frac{\left[nn-1\right]}{\left\langle \xi_{n}n\right\rangle \left\langle \xi_{n-1}n-1\right\rangle }\oint\left(\frac{\left[\tilde{\lambda}n-1\right]\left\langle \xi_{n-1}\lambda\right\rangle \left\langle \xi_{n}\lambda\right\rangle ^{2}}{\left\langle \lambda n\right\rangle \left\langle \lambda n-1\right\rangle \left\langle \xi_{n}n\right\rangle }+\frac{\left[\tilde{\lambda}n\right]\left\langle \xi_{n}\lambda\right\rangle \left\langle \xi_{n-1}\lambda\right\rangle ^{2}}{\left\langle \lambda n\right\rangle \left\langle \lambda n-1\right\rangle \left\langle \xi_{n-1}n-1\right\rangle }\right).
\]
Inserting this into an $(n-2)$-point correlator of hard vertex operators then gives \eqref{eq:subc} after integrating it around each hard vertex operator and adding up the residues. Similarly, \eqref{grpm} arises from braiding the following soft graviton charges
\begin{equation}
\left[q_{n}^{(-1)},\tilde{q}_{n-1}^{(0)}\right]+\left[q_{n}^{(0)},\tilde{q}_{n-1}^{(-1)}\right],
\label{comm2}
\end{equation}
where charges with a tilde are obtained by complex conjugating charges without a tilde. Using the OPE's in \eqref{eq:ope} and and choosing the reference spinors $(\xi_{n-1},\xi_n)=(\lambda_n,\lambda_{n-1})$, one finds that \eqref{comm2} is given by
\[
\frac{1}{2\pi i\left\langle n-1n\right\rangle \left[nn-1\right]}\oint\frac{\left[\tilde{\lambda}n\right]^{2}\left\langle \lambda n-1\right\rangle ^{2}}{\left\langle \lambda n\right\rangle ^{2}\left[\tilde{\lambda}n-1\right]^{2}}\left[\tilde{\lambda}\right|p_{n-1}-p_{n}\left|\lambda\right\rangle, 
\]
which once again gives rise to \eqref{grpm} when inserted into an $(n-2)$-point correlator of hard vertex operators.

\subsection{QED}

The commutator of two soft photon limits can also be understood from the point of view of CFT using Figure \ref{gr2}. Looking at the soft photon theorems in \eqref{qes}, we immediately see that after discarding boundary terms, soft limits involving photons with the same helicity commute, so let us focus on the case where they have opposite helicity. In particular, let particle $n$ have positive helicity and particle $n-1$ have negative helicity and consider the commutator of their soft limits. It is not difficult to see that the leading contribution to the commutator will vanish, so we will focus on terms which contain the leading order soft limit for one particle and the subleading order soft limit for the other. 

Taking particle $n-1$ soft followed by particle $n$ soft gives 
\[
\sum_{i=1}^{n-2}\frac{\left\langle i\xi_{n}\right\rangle }{\left\langle in\right\rangle \left\langle n\xi_{n}\right\rangle }\sum_{j=1}^{n-2}\frac{1}{\left[jn-1\right]}\lambda_{n-1}\cdot\frac{\partial}{\partial\lambda_{j}}\mathcal{A}_{n-2}+\sum_{i=1}^{n-2}\frac{1}{\left\langle in\right\rangle }\tilde{\lambda}_{n}\cdot\frac{\partial}{\partial\tilde{\lambda}_{i}}\left(\sum_{j=1}^{n-2}\frac{\left[j\xi_{n-1}\right]}{\left[jn-1\right]\left[n-1\xi_{n-1}\right]}\mathcal{A}_{n-2}\right),
\]
while taking the soft limits in the reverse order gives:
\[
\sum_{i=1}^{n-2}\frac{1}{\left[jn-1\right]}\lambda_{n-1}\cdot\frac{\partial}{\partial\lambda_{j}}\left(\sum_{i=1}^{n-2}\frac{\left\langle i\xi_{n}\right\rangle }{\left\langle in\right\rangle \left\langle n\xi_{n}\right\rangle }\mathcal{A}_{n-2}\right)+\sum_{j=1}^{n-2}\frac{\left[j\xi_{n}\right]}{\left[jn-1\right]\left[n-1\xi_{n-1}\right]}\sum_{i=1}^{n-2}\frac{1}{\left\langle in\right\rangle }\tilde{\lambda}_{n}\cdot\frac{\partial}{\partial\tilde{\lambda}_{i}}\mathcal{A}_{n-2}.
\] 
After some simplification, one finds that the subleading contribution to the commutator of soft limits is
\begin{equation}
\left[\lim_{p_{n}^{+}\rightarrow0},\lim_{p_{n-1}^{-}\rightarrow0}\right]\mathcal{A}_{n}|_{bulk}=\sum_{i=1}^{n-2}\left(\frac{\left\langle n-1i\right\rangle }{\left\langle in\right\rangle ^{2}\left[in-1\right]}-\frac{\left[ni\right]}{\left\langle in-1\right\rangle ^{2}\left[in\right]}\right)\mathcal{A}_{n-2}.
\label{eq:ymc}
\end{equation}

Equation \eqref{eq:ymc} can be derived from braiding soft photon charges in \eqref{eq:qym}: 
\[
\left[q_{n}^{(-1)},\tilde{q}_{n-1}^{(0)}\right]+\left[q_{n}^{(0)},\tilde{q}_{n-1}^{(-1)}\right],
\]
where

\[
q_{n}^{(-1)}=\frac{1}{2\pi i}\oint\frac{\left\langle \xi_{n}\lambda\right\rangle }{\left\langle n\lambda\right\rangle \left\langle \xi_{n}n\right\rangle }j,\,\,\, q_{n}^{(0)}=\frac{1}{2\pi i}\oint\frac{\left[n\mu\right]}{\left\langle n\lambda\right\rangle }j
\]
\[
\tilde{q}_{n-1}^{(-1)}=\frac{1}{2\pi i}\oint\frac{\left[\xi_{n-1}\tilde{\lambda}\right]}{\left[n-1\tilde{\lambda}\right]\left[\xi_{n-1}n-1\right]}j,\,\,\, \tilde{q}_{n-1}^{(0)}=\frac{1}{2\pi i}\oint\frac{\left\langle n-1\tilde{\mu}\right\rangle }{\left[n-1\tilde{\lambda}\right]}j.
\]
Using the OPE's in \eqref{eq:ope} and neglecting the contribution to the OPE from the current algebra (since photons are abelian), we find that
\[
\left[q_{n}^{(-1)},\tilde{q}_{n-1}^{(0)}\right]+\left[q_{n}^{(0)},\tilde{q}_{n-1}^{(-1)}\right]=\frac{1}{2\pi i}\oint\left(\frac{\left\langle n-1\lambda\right\rangle }{\left[\tilde{\lambda}n-1\right]\left\langle n\lambda\right\rangle ^{2}}-\frac{\left[n\tilde{\lambda}\right]}{\left\langle \lambda n\right\rangle \left[n-1\tilde{\lambda}\right]^{2}}\right).
\]
Plugging this charge into an $(n-2)$-point correlator of hard vertex operators indeed gives \eqref{eq:ymc} after integrating it around each hard vertex operator and adding up the residues.

\end{document}